# Quantum Pseudo-fractional Fourier Transform and its application to quantum phase estimation

Srinivas V. Parasa[1], K. Eswaran[2]
[1]Dept. of Electrical and Computer Engineering, Portland State University, Portland, Oregon, USA
[2]Dept. of Computer Science, Sree Nidhi Institute of Science & Technology(JNTU), Hyderabad, A.P., India

**Abstract -** *In this paper we present a method to compute the coefficients of the fractional Fourier transform (FrFT)[1] on a quantum computer using quantum gates of polynomial complexity of the order $O(n^3)$. The FrFt, a generalization of the DFT, has wide applications in signal processing and is particularly useful to implement the Pseudopolar[3] and Radon transforms [3]. Even though the FrFT is a non-unitary operation, to develop its quantum counterpart, we develop a unitary operator called the quantum Pseudo-fraction Fourier Transform (QPFrFT) in a higher-dimensional Hilbert space, in order to computer the coefficients of the FrFT. In this process we develop a unitary operator denoted by $U_\alpha$ which is an essential step to implement the QPFrFT. We then show the application of the operator $U_\alpha$ in the problem of quantum phase estimation.*

**Keywords:** Fractional Fourier transform (FFT), Quantum Pseudo-fractional Fourier transform (QPFrFT), Quantum phase estimation.

## 1 Introduction

In this paper, we present a method to compute the linear[1] fractional Fourier transform coefficients on a quantum computer and show how it can be used in the problem of quantum phase estimation. The organization of the paper is as follows: In section 2, we present a brief introduction to the the classical FrFT and then in section 3, we proceed to develop a quantum version of the FrFT, also called as Quantum Pseudo-fractional Fourier transform (QPFrFT) using a novel quantum operator developed in this paper, which is denoted as $U_\alpha$. In section 4, we develop the tensor product notation useful to implement the $U_\alpha$ operator (and thus the QPFrFT) and then show in section 5, the procedure to implement a quantum circuit of polynomial gate complexity to implement the $U_\alpha$ operator. Finally, in section 6, we show, using three different methods, as to how the $U_\alpha$ operator can be used for the problem of phase estimation and then conclude our paper in section 7 with possible future applications of the $U_\alpha$ operator and improvements to our present work.

## 2 Fractional Fourier transform

The linear fractional Fourier transform [1] is a special case of Chirp-Z transform [1]. The discrete fractional Fourier transform on a 1-D discrete signal say $f[j]$, $0 \leq j < N$ and given an arbitrary scalar $\alpha$, is denoted by $F_\alpha[k]$ and is defined as follows:

$$F_\alpha[k] = \sum_{j=0}^{N-1} f[j] \cdot \exp(2\pi i \cdot \frac{kj}{N} \cdot \alpha) \qquad (1)$$

When $\alpha = 1$, equation (1) reduces to the Fourier transform and when $\alpha = -1$ it reduces to the Inverse Fourier transform. Thus the fractional Fourier transform is a very general definition. The 1D fractional Fourier transform can be evaluated in $30N \log N$ operations [3].

*Implementation of the classical FrFT:* Making the substitution $-2kj = k^2 + j^2 - (k-j)^2$ in the exponential in equation (1) and after simplification, we obtain the following equation

$$F_\alpha[k] = s(k) \cdot \{(f(j) \cdot s(j)) \otimes s(j)\}, where\ s(j) = e^{(-i \cdot \frac{\pi \cdot \alpha}{N} \cdot j^2)} \qquad (2)$$

In equation (2) the symbol $\otimes$ indicates convolution. (A word of caution: in the subsequent sections $\otimes$ denotes a direct product in a Hilbert space, this clash of notation can be easily distinguished from the context). Hence it is easy to see that the 1D fractional Fourier transform can be evaluated in $30N \log N$ operations as a sequence of 1D FFT operations [3]

## 3 Quantum version of the discrete FrFT

Any attempts to implement the FrFT on the quantum computer on the lines of classical techniques have to abandoned because i) FrFT is not a unitary operation ii) quantum convolution is not possible [5]. However, in this paper we will present a method to *compute the coefficients of the FrFT* using 'quantum' methods.

---
[1] It must be noted that, in the literature there are two completely different mathematical tools and both of these are misleadingly called by the name "fractional Fourier transform". The first kind is known as the *linear* fractional Fourier transform [1,3] while the second one is known as the *quadratic* fractional Fourier transform [4] whose quantum version was developed in [6,7]. *We do not use the quadratic FrFT anywhere else in this paper. Henceforth, we refer to the Linear FrFT as FrFT, omitting the word linear.*

## 3.1 Quantum Approach to FrFt

We want to implement the fractional Fourier transform denoted by $F_\alpha(k)$, which is not a unitary transformation except for $\alpha = \pm 1$. Therefore, by conventional wisdom; it is not at all possible to implement the fractional Fourier transform on a quantum computer. In this section, we overcome the 'unitarity' constraint and show that, it is indeed possible to compute the fractional Fourier transform *coefficients* $F_\alpha(k)$ using unitary operations in an enlarged Hilbert space. The technique is as follows: instead of trying to realize the state $\sum_{k=1}^{N-1} F_\alpha(k)|k\rangle$ starting from an initial state $\sum_{j=1}^{N-1} f(j)|j\rangle$ to compute the coefficients $F_\alpha(k)$, it suffices to implement a unitary operator $U_{PF_\alpha}$ in an enlarged Hilbert space say $|Q\rangle \otimes |k\rangle$ to realize the transformation given below;

$$\sum_{j=1}^{N-1} f(j)|j\rangle|0\rangle^n \xrightarrow{U_{PF_\alpha}} \sum_{Q=0,k=0}^{N-1,N-1} F(Q,k)|Q\rangle \otimes |k\rangle \quad (3)$$

Such that $F(0,k) = F_\alpha(k)$. Let the unitary operator $U_{PF_\alpha}$ be referred to as quantum pseudo-Fractional Fourier transform (QPFrFT). Before we present the QPFrFT, we present and define another unitary operator $U_\alpha$ which is essential to implement the operator $U_{PF_\alpha}$.

## 3.2 Operator $U_\alpha$

We define $\alpha = \dfrac{l - 2^n}{2^n}$ $(0 \le l < 2^{n+1})$. Thus $\alpha$ can be indirectly represented by the state vector $|l\rangle$. We then define a controlled unitary operator $U_\alpha$ which implements the following unitary operation in the enlarged Hilbert space defined by $|l\rangle \otimes |j\rangle \otimes |0\rangle^n$ (where $|0\rangle^n = \bigotimes_{n=1}^{n} |0\rangle$) which is also shown symbolically in fig (1).

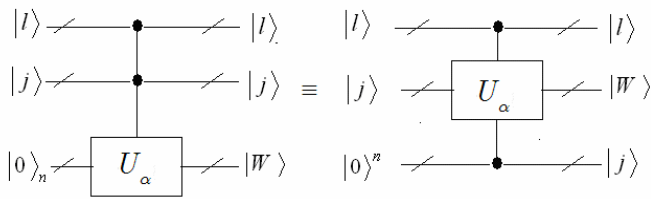

Fig (1) Quantum circuit representing the controlled gate $U_\alpha$

We define $U_\alpha$ as an operator which performs the following:

$$|l\rangle|j\rangle|0\rangle^n \xrightarrow{U_\alpha} |l\rangle \otimes |j\rangle \otimes |W\rangle$$
$$\text{Where } |W\rangle = \left(\frac{1}{\sqrt{2^n}} \sum_{k=0}^{N-1} \exp(i \cdot 2\pi \cdot \frac{\alpha k j}{2^n}) \, |k\rangle\right) \quad (4)$$

The unitarity of the operator $U_\alpha$ can be easily verified.

## 3.3 Quantum Pseudo-fractional Fourier transform (QPFrFT)

Now, coming to the problem of computation of the fractional Fourier transform coefficients, for a given $\alpha$, the state vector $|l\rangle$ is fixed. If we start with a superposition input state $|l\rangle \left(\sum_{j=1}^{N-1} f(j)|j\rangle\right)|0\rangle^n$ and perform the following operation:

$$|l\rangle \left(\sum_{j=0}^{N-1} f(j)|j\rangle\right)|0\rangle^n \xrightarrow{U_\alpha} |l\rangle \otimes \left(\sum_{j=0}^{N-1} f(j)|j\rangle \otimes |W\rangle\right)$$
$$= \left(|l\rangle \otimes \sum_{j=0}^{N-1} \left[f(j)|j\rangle \otimes \left(\frac{1}{\sqrt{2^n}} \sum_{k=0}^{N-1} \exp(i \cdot 2\pi \cdot \frac{\alpha k j}{2^n}) \, |k\rangle\right)\right]\right) \quad (5)$$
$$= \left(|l\rangle \otimes \frac{1}{\sqrt{2^n}} \sum_{j=0}^{N-1} \sum_{k=0}^{N-1} \left(f(j) \cdot \exp(i \cdot 2\pi \cdot \frac{\alpha k j}{2^n})|j\rangle \otimes |k\rangle\right)\right)$$

Rearranging the order of summations, in the above expression and using the notation given below

$$X_\alpha(j,k) = \frac{1}{\sqrt{2^n}} \cdot f(j) \cdot \exp(i \cdot 2\pi \cdot \frac{\alpha k j}{2^n}) \quad (6)$$

we have the following output state.

$$|l\rangle \left(\sum_{j=0}^{N-1} f(j)|j\rangle\right)|0\rangle^n \xrightarrow{U_\alpha} |l\rangle \sum_{k,j=0}^{N-1,N-1} X_\alpha(j,k) \bigl(|j\rangle \otimes |k\rangle\bigr) \quad (7)$$

And also from equation (1) we note that FrFt coefficients denoted by $F_\alpha(k) = \sum_{j=0}^{N-1} X_\alpha(j,k)$. Thus the quantum circuit $U_\alpha$ calculates all the coefficient terms required to calculate the fractional Fourier transform coefficients $F_\alpha(k)$.

Now we have to sum up the coefficients $X_\alpha(j,k)$ to obtain $F_\alpha(k)$. We implement this by applying a QFT on the state vector $|j\rangle$. (We know that after applying a FT, the first coefficient in the spectral domain is nothing but the sum of all the values of the original function). Applying a QFT on $|j\rangle$ i.e. the unitary operation $I \otimes F \otimes I$ on $|l\rangle|j\rangle|k\rangle$ gives the following:

$$|l\rangle \sum_{k=0}^{N-1}\left(\sum_{j=0}^{N-1} X_\alpha(j,k)|j\rangle\right)|k\rangle \xrightarrow{I \otimes F \otimes I} |l\rangle \sum_{k=0}^{N-1}\left(\sum_{Q=0}^{N-1} Z_\alpha(Q,k)|Q\rangle\right)|k\rangle$$

Where $Z_\alpha(Q,k)$ is the 1D FT of the function $X_\alpha(j,k)$ on $j$. Then, by definition we can write the values of $Z_\alpha(0,k)$ as:

$$Z_\alpha(Q,k) = \frac{1}{\sqrt{N}} \sum_{j=0}^{N-1}\left(X_\alpha(k,j) \times e^{\frac{2\pi i}{N} \times Q \times j}\right)$$
$$Z_\alpha(0,k) = \frac{1}{\sqrt{N}} \sum_{j=0}^{N-1} X_\alpha(j,k) = \frac{F_\alpha(k)}{\sqrt{N}} \quad (8)$$

Thus we have computed all the coefficients of the FrFt. Thus the amplitude of state $|l\rangle|0\rangle|k\rangle$ or simply $|0\rangle \otimes |k\rangle$ (as $|l\rangle$ is fixed vector for a given $\alpha$) indirectly gives the value $F_\alpha(k)$ $(=\sqrt{N} \cdot Z_\alpha(0,k))$ which is nothing but the fractional Fourier transform we wanted to compute.

For convenience, we shall refer to the above transform as *quantum Pseudo-fractional Fourier transform* and denote it by $U_{PF_\alpha}$. The implementation of the quantum Pseudo fractional Fourier transform is illustrated pictorially in fig (3).

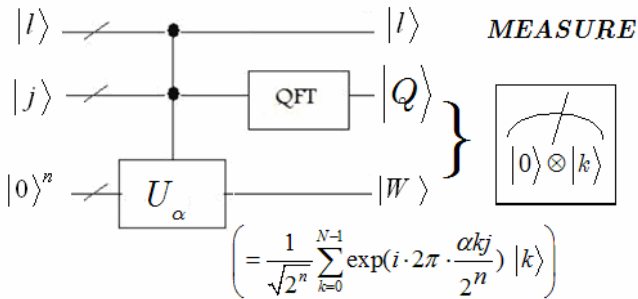

Figure (2) Quantum Pseudo-fractional Fourier transform (QPFrFT)

## 4  Tensor product notation useful to implement the $U_\alpha$ Operator

Consider the ket $|\Omega\rangle = \frac{1}{\sqrt{2^n}} \sum_{k=0}^{N-1} \exp(i \cdot \frac{2\pi kj}{2^n} \cdot \alpha) |k\rangle$. Similar to the QFT [8], it is easy to develop a tensor product notation for the Quantum Pseudo-fractional FFT operation.

*Truncation of multiples of $2\pi i$*: In the exponent of the ket $|\Omega\rangle$, only the terms in the expression $\frac{\alpha kj}{2^n}$ which are binary fractions remain while the other terms are truncated (as they are multiples of $2\pi i$).

Using the notation $k = \sum_{p=1}^{n} k_p \cdot 2^{n-p}, j = \sum_{r=1}^{n} j_r \cdot 2^{n-r}$ and $l = \sum_{q=1}^{n+1} l_q 2^{(n+1-q)} = l_1 2^n + \sum_{q=1}^{n} l'_q 2^{(n-q)}$, when substituted in $\frac{\alpha kj}{2^n}$ and after the collection of all the binary fractions, we get the following.

$$\frac{\alpha kj}{2^n} = \sum_{p=1}^{n} k_p \underbrace{\left((l_1-1) \sum_{r=n-p+1,r>0}^{n} j_r \cdot 2^{(n-r-p)}\right)}_{\Phi_p} + \\ \sum_{p=1}^{n} k_p \underbrace{\left(\sum_{q=1}^{n} l'_q \sum_{r=(n+1)-(p+q),r>0}^{n}\left(j_r \cdot 2^{(n-r)-(p+q)}\right)\right)}_{\Theta_p} \quad (9)$$

$$\frac{\alpha kj}{2^n} = \sum_{p=1}^{n} k_p \Phi_p + \sum_{p=1}^{n} k_p \Theta_p = \sum_{p=1}^{n} k_p \underbrace{\left(\Phi_p + \Theta_p\right)}_{\Psi_p} = \sum_{p=1}^{n} k_p \Psi_p \quad (10)$$

### 4.1  Nature of $\Phi_p$ and $\Theta_p$

After expanding the terms in $\Phi_p$ and $\Theta_p$, we get the following expressions:

$$\Theta_n = l'_1(0.0 j_1 j_2 j_3 ... j_n) + l'_2(0.00 j_1 j_2 j_3 ... j_{n-1} j_n) + .... \\ + l'_{n-1}(0.\underbrace{00...00}_{(n-1\ zeros)} j_1 j_2 j_3 ... j_n) + l'_n(0.\underbrace{00...00}_{(n\ zeros)} j_1 j_2 j_3 ... j_n)$$

$$\Theta_{n-1} = l'_1(0. j_1 j_2 j_3 ... j_n) + l'_2(0.0 j_1 j_2 j_3 ... j_{n-1} j_n) + ..... \\ + l'_{n-1}(0.\underbrace{00...00}_{(n-2\ zeros)} j_1 j_2 j_3 ... j_n) + l'_n(0.\underbrace{00...00}_{(n-1\ zeros)} j_1 j_2 j_3 ... j_n)$$

$$\Theta_2 = l'_1(0. j_{n-2} j_{n-1} j_n) + l'_2(0. j_{n-3} j_{n-2} j_{n-1} j_n) + ... \\ + l'_{n-1}(0.0 j_1 j_2 j_3 ... j_n) + l'_n(0.00 j_1 j_2 j_3 ... j_n)$$

$$\Theta_1 = l'_1(0. j_{n-1} j_n) + l'_2(0. j_{n-2} j_{n-1} j_n) + .... \\ + l'_{n-1}(0. j_1 j_2 j_3 ... j_n) + l'_n(0.0 j_1 j_2 j_3 ... j_n) \quad (11\text{-}14)$$

Also $\Phi_p = (l_1-1) \sum_{r=n-p+1,r>0}^{n} j_r \cdot 2^{(n-r-p)}$ and hence $\Phi_p$ solely depends on $l_1$.

## 4.2 Tensor product

Thus replacing $\frac{\alpha kj}{2^n} = \sum_{p=1}^{n} k_p \Psi_p$ in the exponent of the expression for the original ket $|\Omega\rangle$, we get a tensor product notation similar to the QFT, as shown below.

$$\frac{1}{\sqrt{2^n}}\sum_{k=0}^{N-1}\exp(i\cdot 2\pi\cdot\frac{\alpha kj}{2^n})\,|k\rangle = \frac{1}{\sqrt{2^n}}\sum_{k=0}^{N-1}\exp(i\cdot 2\pi\cdot[\sum_{p=1}^{n}k_p\Psi_p])|k\rangle$$

$$= \frac{1}{\sqrt{2^n}}\left\{\bigotimes_{p=1}^{n}\left[\sum_{k_p=0}^{1}\exp(i\cdot 2\pi\cdot k_p\Psi_p)|k_p\rangle\right]\right\}$$

$$= \left\{\bigotimes_{p=1}^{n}\left(\frac{[|0\rangle+\exp(i\cdot 2\pi\cdot\Psi_p)|1\rangle]}{\sqrt{2}}\right)\right\} \quad (15)$$

Let us denote, $|W_p\rangle = \frac{|0\rangle+e^{i\cdot 2\pi\cdot\Psi_p}|1\rangle}{\sqrt{2}}$. Hence, we can write

$$|\Omega\rangle = \frac{1}{\sqrt{2^n}}\sum_{k=0}^{N-1}\exp(i\cdot 2\pi\cdot\frac{\alpha kj}{2^n})\,|k\rangle = \bigotimes_{p=1}^{n}|W_p\rangle \quad (16)$$

## 5 Quantum circuit to implement $U_\alpha$

Similar to the QFT, because of the tensor product notation, it is easy to implement the QPFrFT using single qubit controlled rotation gates [11]. Due to the large number of terms involved and since the procedure is same for any value of $N$, we wish to briefly illustrate the procedure to implement the quantum circuit for the operator $U_\alpha$. For the case of $N=4$, we have

$$|l_1\rangle|l_2\rangle|l_3\rangle|j_1\rangle|j_2\rangle|0\rangle_1|0\rangle_2 = |l_1\rangle|l_1'\rangle|l_2'\rangle|j_1\rangle|j_2\rangle|0\rangle_1|0\rangle_2$$

$$\xrightarrow{U_\alpha}|l\rangle|j\rangle\frac{(|0\rangle+e^{2\pi i\Psi_1}|1\rangle)}{\sqrt{2}}\frac{(|0\rangle+e^{2\pi i\Psi_2}|1\rangle)}{\sqrt{2}} \quad (17)$$

Where, $\Psi_p = \Theta_p + \Phi_p (1\le p\le 2)$. We then have to implement the following terms:

$$\Theta_1 = l_1'(\frac{j_1}{2}+\frac{j_2}{2^2})+l_2'(\frac{j_1}{2^2}+\frac{j_2}{2^3}),\ \Phi_1 = (-1)\overline{l_1}\cdot\frac{j_1}{2} \quad (18)$$

$$\Theta_2 = l_1'(\frac{j_1}{2^2}+\frac{j_2}{2^3})+l_2'(\frac{j_1}{2^3}+\frac{j_2}{2^4}),\ \Phi_2 = (-1)\overline{l_1}(\frac{j_1}{2}+\frac{j_2}{2^2}) \quad (19)$$

We now show a general method to perform the transformation denoted by $|0\rangle_p \to \frac{|0\rangle+e^{2\pi i\Psi_p}|1\rangle}{\sqrt{2}}$

1) Applying the Hadamard gate $H$ on $|0\rangle_1$ as shown in figure (3a) gives $|+\rangle = \left(\frac{|0\rangle+|1\rangle}{\sqrt{2}}\right)$.

2) Let us define the controlled rotation gates $\overset{a,b}{R_k} = \begin{pmatrix} 1 & 0 \\ 0 & e^{2\pi i\cdot ab/2^k} \end{pmatrix}$ and $\overset{a,b}{\overline{R}_k} = \begin{pmatrix} 1 & 0 \\ 0 & e^{-2\pi i\cdot ab/2^k} \end{pmatrix}$ as the rotation gates with control qubit $|a\rangle$ and target qubit $|b\rangle$. After applying the cascade of controlled rotation gates $R_1^{l_1',j_1}\times R_2^{l_1',j_2}\times R_2^{l_2',j_1}\times R_3^{l_2',j_2}$ on the state $|+\rangle$ we get the state shown below (also illustrated in figure (3a)).

$$\frac{|0\rangle+e^{i\cdot 2\pi\left(l_1'\left[\frac{j_1}{2}+\frac{j_2}{2^2}\right]+l_2'\left[\frac{j_1}{2^2}+\frac{j_2}{2^3}\right]\right)}|1\rangle}{\sqrt{2}} = \frac{|0\rangle+e^{i\cdot 2\pi\cdot\Theta_1}|1\rangle}{\sqrt{2}} \quad (20)$$

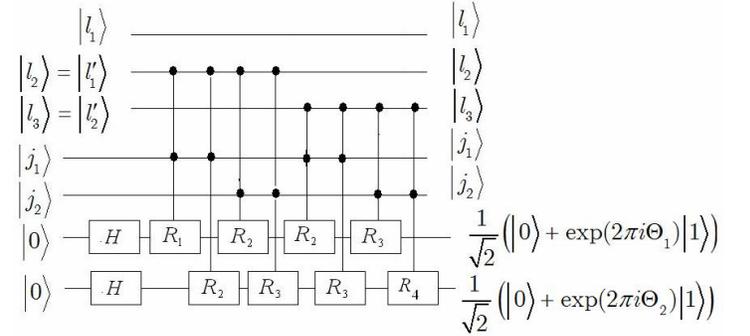

Figure (3a)

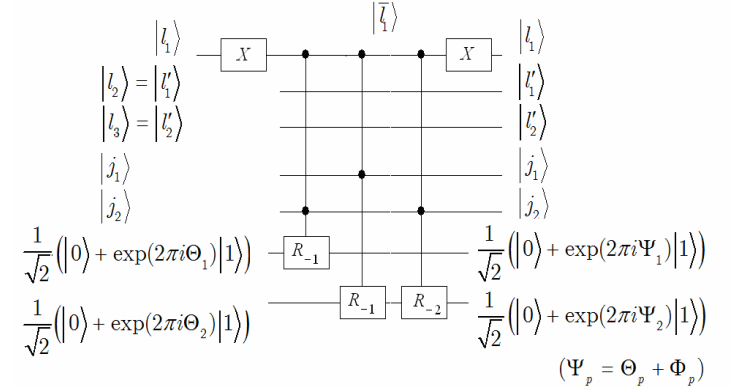

Figure (3b)

Figure (3) Quantum circuit to implement the $U_\alpha$ operator

3) Now applying Pauli $X$ gate on $|l_1\rangle$ produces $|\overline{l_1}\rangle$.

$$\frac{|0\rangle+e^{i\cdot 2\pi\cdot\Theta_1}|1\rangle}{\sqrt{2}}\xrightarrow{\overline{R}_1^{\overline{l_1},j_1}}\frac{|0\rangle+e^{-i\cdot 2\pi\left(\overline{l_1}\frac{j_1}{2}\right)}e^{i\cdot 2\pi\cdot\Theta_1}|1\rangle}{\sqrt{2}}$$

$$= \frac{|0\rangle+e^{i\cdot 2\pi\cdot\Phi_1}e^{i\cdot 2\pi\cdot\Theta_1}|1\rangle}{\sqrt{2}} = \frac{|0\rangle+e^{i\cdot 2\pi\cdot\Psi_1}|1\rangle}{\sqrt{2}} \quad (21)$$

(Illustrated in figure (3b). In the figure (3), the gates $\overset{a,b}{\overline{R}_k}$ is represented as $\overset{a,b}{R_{-k}}$). Finally we apply Pauli $X$ gate on $|\overline{l_1}\rangle$ to get back $|l_1\rangle$. Similarly the transformation $|0\rangle_p \to \frac{|0\rangle + e^{2\pi i \Psi_p}|1\rangle}{\sqrt{2}}$ can be obtained for all values of $p$ by applying a cascade of appropriate combinations of the controlled $R_k^{a,b}$ and $\overline{R}_k^{a,b}$ rotation gates (shown in figure (3)). The above technique can be easily generalized to any $N$.

## 5.1 Calculation of the gate count

It is easy to see that, for an arbitrary $N$, we need a total of $n^2(n+1)$ $R_k^{a,b}$ gates to implement the term $e^{i \cdot 2\pi \cdot \Theta_p}$ and another $n(n+1)/2$ $\overline{R}_k^{a,b}$ gates to implement the term $e^{i \cdot 2\pi \cdot \Phi_p}$. Using techniques developed in [9,10] to implement controlled gates using single qubit and two-qubit operations, we obtain the total gate of the order $O(n^3)$ which is comparable to that of Shor's quantum Fourier transform ($O(n^2)$).

## 6 Phase Estimation using $U_\alpha$ Operator

Let $|u\rangle$ be the eigen state of unitary operator $U$ with eigen value $e^{2\pi i \varphi_u}$. We now wish to show $U_\alpha$ operator based techniques to determine the best approximation to the phase $\varphi_u$. In the following sections, we describe three methods of using $U_\alpha$ for phase estimation.

## 6.1 $U_\alpha$ based phase-estimation using lesser number of $U^k$ gates.

The steps involved in the phase estimation algorithm we developed are as follows:

*STEP 1:* We start with a superposition state given below and do the following transformation.

$$|l\rangle|j\rangle|0\rangle^n|u\rangle \xrightarrow{U_\alpha} |l\rangle|j\rangle\left(\frac{1}{\sqrt{N}}\sum_{k=0}^{N-1}\exp(i \cdot 2\pi \cdot \frac{\alpha_l k j}{2^n})|k\rangle\right)|u\rangle$$

We now define a controlled gate $U^k$ as follows: a $U^k$ gate is a controlled gate with its control qubit being $|k\rangle$. Thus $U^k$ gate implies that the operator $U$ is applied $k$ times on the target qubit. *Note:* $\alpha_l = \frac{l-L}{L}$ ($0 \le l < 2L, L = 2^l$)

*STEP 2:* Now we apply the controlled gates $U^k$ to get

$$|l\rangle|j\rangle\left(\frac{1}{\sqrt{N}}\sum_{k=0}^{N-1}\exp(i \cdot 2\pi \cdot \frac{\alpha_l k j}{2^n}) \times \exp(i \cdot 2\pi \cdot k \cdot \varphi_u)|k\rangle\right)|u\rangle$$

In our method we need to apply fewer $U^k$ gates compared to the standard phase estimation procedure [11] to do phase estimation.

*STEP 3:* Re arranging the above terms we get

$$|l\rangle|j\rangle\left(\frac{1}{\sqrt{N}}\sum_{k=0}^{N-1}\exp\left(\frac{i \cdot 2\pi \cdot k}{N}(\alpha_l j + N\varphi_u)\right)|k\rangle\right)|u\rangle \quad (22)$$

Let $\varphi_u \approx \tilde{\varphi}_u$ be represented up to $n'$ bits of precision.

Then $\varphi_u \approx \tilde{\varphi}_u = \frac{b}{2^{n'}} = \frac{b}{N'} = 0.\tilde{\varphi}_1 \tilde{\varphi}_2 \tilde{\varphi}_3 \ldots \tilde{\varphi}_{n'-1} \tilde{\varphi}_{n'}$. Thus, if we use the standard phase estimation algorithm presented in standard quantum computing textbooks (for ex. [11]) to obtain $\varphi_u \approx \tilde{\varphi}_u$ up to $n'$ bits of precision with probability of success at least $1-\varepsilon$, from [11], we require a total number of $q$ qubits, where $q$ is given by $q = n' + \log\left(2 + \frac{1}{2\varepsilon}\right)$.

This implies that we should be able to implement $U^k$ gates where $k$ is of the form $k = 2^i$ ($0 \le i \le q-1$). It has been reported that the physical implementation of the higher order $U^k$ is quite difficult [12].

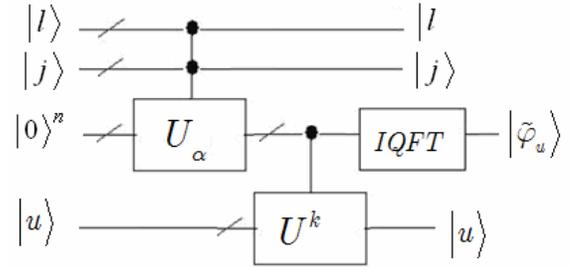

Figure (4) Illustration of the $U_\alpha$ based phase estimation

Now we show a method to perform phase estimation using lesser number of higher order $U^k$ gates. In our case $k = 2^i$ ($0 \le i \le n-1$) and $n < n' < q$.

Consider the exponential $\exp\left(\frac{i \cdot 2\pi \cdot k}{N}(\alpha_l j + N\varphi_u)\right)$. Let $\phi = (\alpha_l j + N\varphi_u)$. After making the substitution, the state in (22) can be written as

$$|l\rangle|j\rangle\left(\frac{1}{\sqrt{N}}\sum_{k=0}^{N-1}\exp\left(\frac{i \cdot 2\pi \cdot k}{N}\phi\right)|k\rangle |u\rangle\right) \quad (23)$$

*STEP 4:* Now after, taking the IQFT on qubit $|k\rangle$, we get

$$|l\rangle|j\rangle\left(\frac{1}{\sqrt{N}}\sum_{k=0}^{N-1}\exp^{\left(\frac{i \cdot 2\pi \cdot k}{N}\phi\right)}|k\rangle |u\rangle\right) \xrightarrow{IQFT\ on\ k} |l\rangle|j\rangle|\tilde{\phi}\rangle|u\rangle \quad (24)$$

Thus we can estimate $\tilde{\phi}$ and thus $\tilde{\varphi}_u$. In this method there is a trade-off between the precision of the estimated phase and

## 6.2 Phase estimation as a Grover search

While the algorithm presented in section 6.1 is more or less on the lines of the standard phase estimation, with the added advantage of reducing the number of higher powers of the $U$ gates, in this section, we present a method where we generate a lot of phase terms using the $U_\alpha$ and try to match one of those terms with the phase we actually want to estimate. This is performed by doing a Grover search using the method below.

In the STEP 1 of the previous algorithm in section 6.1, if we start with a superposition state with different values of $l$, i.e. $\left(\frac{1}{\sqrt{L}}\sum_{l=0}^{L-1}|l\rangle|j\rangle|0\rangle^n\right)|u\rangle$ and after performing STEPS 1-3, we will obtain the following state.

$$\frac{1}{\sqrt{L}}\left(\sum_{l=0}^{L-1}\left(\frac{1}{\sqrt{N}}\sum_{k=0}^{N-1}\exp\left(\frac{i\cdot 2\pi\cdot k}{N}(\alpha_l j+N\varphi_u)\right)|l\rangle|j\rangle|k\rangle|u\rangle\right)\right) \quad (25)$$

Replace $(\alpha_l j+N\varphi_u)=\overbrace{\left(-\frac{l}{L}\right)}^{\alpha_l}\cdot j+N\cdot\overbrace{\left(\frac{b}{N'}\right)}^{\varphi_u}=\phi$ in (25).

$$|l\rangle|j\rangle\left(\frac{1}{\sqrt{N}}\sum_{k=0}^{N-1}\exp\left(\frac{i\cdot 2\pi\cdot k}{N}\phi\right)|k\rangle|u\rangle\right) \quad (26)$$

The main approach is to perform a search to find that value of $l$ for which $\phi=0$. For this, we can also impose the following conditions 1) $\frac{j}{L}=\frac{N}{N'}$ and 2) $N'\leq L$ which gives $\phi=\frac{N}{N'}(b-l)$ and proceed as follows:

We now apply IQFT on the qubit $|k\rangle$ in the state (24) to obtain $\frac{1}{\sqrt{L}}\sum_{l=0,\tilde\phi=0}^{L,N}|l\rangle|j\rangle|\tilde\phi\rangle|u\rangle \quad (27)$

We do a Grover search for $|\tilde\phi\rangle=0$ and obtain the estimate of the phase, $\tilde\varphi_u=\frac{b}{N'}\left(-\frac{\alpha_l\cdot j}{N}=\frac{l}{L}\cdot\frac{j}{N}\right)$ with a high probability. Thus using both the above methods we have reduced the number of higher order $U^k$ gates. It must be noted that the limitation of this method is that it gives correct results only when the phase $\varphi_u=\frac{b}{2^n}$, i.e. an exact binary fraction. Nevertheless, the novel aspect of this method is that it shows that one can use Grover's algorithm for the phase estimation.

Using Grover search appears to give an impression that it slows the QPE algorithm from a complexity of $O(n'^2)$ using the standard method to $O(\sqrt{N})$. We show that this is not necessarily true.

For the Grover based QPE algorithm to achieve the same or better performance as that of the traditional QPE algorithm, we need to have the condition (Ignoring the associated constants) $O(\sqrt{N})\leq O\left((n')^2\right)\Rightarrow N\leq(n')^4$. We also have $N<N'$. Let $\frac{N'}{N}=2^v$. Substituting this in the above inequality, and after some algebraic manipulation we get:

$$2^n\leq(v+n)^4 \quad (28)$$

We can always improve the accuracy of the estimation the phase $\varphi_u=\frac{b}{2^{n'}}$ by having a large $N'$, which means a large $v$. We can thus ensure that given an $n$, we can always find a large $v$ (hence large $N'$) such that the inequality (28) is always satisfied. For example, if $v=0$, the inequality (28) is always satisfied for $n<16$. Thus we can estimate the phase up to an accuracy of $\frac{1}{2^{16}}\approx 10^{-5}$ i.e. we can get a correct result up to a precision of 5 decimal places using the Grover's method, which is quite reasonable for many applications. As another example if $v=16$, the inequality (28) is always satisfied for the value of $n<21$ as shown in figure (5).

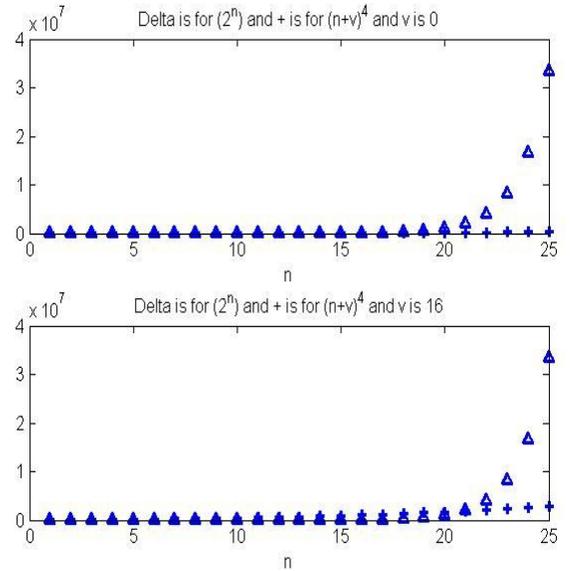

Figure (5) Solution of the inequality $2^n\leq(v+n)^4$ (28) for $v=0$ (top), 16 (bottom). We are plotting $2^n$ using 'delta' symbol and $(v+n)^4$ using the '+' symbol.

## 6.3 Using $U_\alpha$ operator to modulate the estimated value of the phase by the eigen state

We now show another method involving the $U_\alpha$ operator to do phase estimation. Let $|u\rangle$ be the eigen state of unitary operator $U$ with eigen value $e^{2\pi i \varphi_u}$.

Let $\varphi_u \approx \dfrac{\tilde{\varphi}_u}{2^n} = 0.\tilde{\varphi}_1 \tilde{\varphi}_2 .. \tilde{\varphi}_{n-1} \tilde{\varphi}_n$ be the best $n$ bit approximation to $\varphi_u$.

STEP 1: We start with the initial state $|l\rangle|u\rangle|0\rangle^n$. After applying $U_\alpha$ operator with control qubits as $|l\rangle, |u\rangle$ and target qubits as $|0\rangle^n$ we get the following state.

$$|l\rangle|u\rangle|0\rangle^n \xrightarrow{U_\alpha} |l\rangle\left(\frac{1}{\sqrt{2^n}}\sum_{k=0}^{N-1} e^{-\left(\frac{2\pi i \alpha k u}{2^n}\right)}|u\rangle|k\rangle\right) \quad (29)$$

STEP 2: Apply controlled gates $U^k$ to get

$$\frac{1}{\sqrt{2^n}}\sum_{k=0}^{N-1} e^{-\left(\frac{2\pi i \alpha k u}{2^n}\right)} (U^k|u\rangle)|k\rangle \to \frac{1}{\sqrt{2^n}}\sum_{k=0}^{N-1} e^{-\left(\frac{2\pi i \alpha k u}{2^n}\right)} e^{2\pi i k \varphi_u}|u\rangle|k\rangle$$

$$= \frac{1}{\sqrt{2^n}}\sum_{k=0}^{N-1} e^{2\pi i k\left(\varphi_u - \left(\frac{\alpha u}{2^n}\right)\right)}|u\rangle|k\rangle \quad (30)$$

STEP 3: Applying Inverse QFT on $|k\rangle$ we obtain $|u\rangle|\tilde{\varphi}_u - u\alpha\rangle$ (we choose $\alpha$ such that $u\alpha$ is an integer). Now we measure the first register to obtain $\tilde{\varphi}_u - u\alpha$ and can easily obtain $\tilde{\varphi}_u$ as illustrated in figure (6).

It is interesting to note that the estimate of the phase value is linearly modulated by its eigen vector. Thus one can tune the parameter $\alpha$ in the state $|u\rangle|\tilde{\varphi}_u - u\alpha\rangle$ to obtain the state $|u\rangle|0\rangle$.

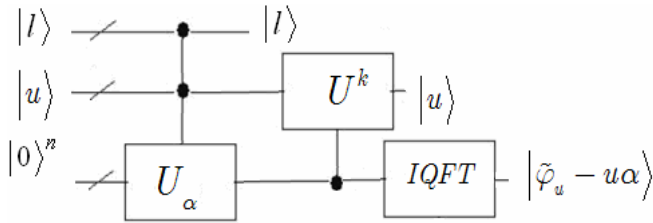

Figure (6) Illustration of $U_\alpha$ based phase estimation

Thus we have shown *three different* variations of the quantum phase estimation algorithm using the $U_\alpha$ operator.

## 7 Conclusion

In this paper, we have shown a method to compute the non-unitary fractional FFT coefficients using the quantum Pseudo-fractional FT by developing the $U_\alpha$ operator and showed how it can be effectively applied to the problem of phase estimation which is a very important subroutine in the problem of order finding and factoring numbers [11].

There are many other applications of the FrFT. It is an essential tool in the calculation of the Pseudopolar transform [3] (PPFFT) and Radon transform [3]. Thus QPFrFt can be used to implement the quantum version of the PPFFT. We have also a developed a method to implement the Quantum PPFFT ($O(n^3)$) [2]. The classical Pseudopolar transform is useful for image registration. Thus the QPPFFT may be applied to the problem of detection of rotations in images similar to the method proposed in [13].